\input harvmac
%
%
%
%
%
 
\parindent=0pt
\Title{\vbox{\baselineskip12pt\hbox{SHEP 97-03}
}}{Large $N$ and the Renormalization Group}

\centerline{\bf Marco D'Attanasio and Tim R. Morris}
\vskip .12in plus .02in
\centerline{\it 
Department of Physics, University of Southampton,}
\centerline{\it Highfield, Southampton SO17 1BJ, UK}
\vskip .7in plus .35in

\centerline{\bf Abstract}
\smallskip 
In the large $N$ limit, we show that
the Local Potential Approximation to the flow equation for
the Legendre effective action,
is in effect no longer an approximation, but
exact -- in a sense, and under conditions, that  we determine precisely.
 We  explain why the same is not true for 
the Polchinski or Wilson flow equations and,
by deriving an exact relation between the Polchinski and Legendre
effective potentials (that holds for all $N$),
we find the correct large
$N$ limit of these flow equations. We also show that all forms 
 (and all parts) of the renormalization
group are exactly soluble in the large $N$ limit, 
choosing as an example, $D$ dimensional
$O(N)$ invariant $N$-component scalar field 
theory.\footnote{$\dagger$}{This footnote
does not appear in the version to be 
published in Phys. Lett. B.}

\vskip -1.5cm
\Date{\vbox{
{hep-th/9704094}
\vskip2pt{April, 1997.}
}
}

\def\ins#1#2#3{\hskip #1cm \hbox{#3}\hskip #2cm}

\def\ie{{\it i.e.}\ }
\def\eg{{\it e.g.}\ }
\def\cf{{\it c.f.}\ }
\def\viz{{\it viz.}\ }
\def\aka{{\it a.k.a.}\ }

\def\nonp{non-perturbative}
\def\phi{\varphi}
\def\D{{\cal D}}

\def\q{{\bf q}}
\def\x{{\bf x}}
\def\y{{\bf y}}


\parindent=15pt

\noindent{\bf 1. Introduction.}

The large $N$ limit\foot{$N$ being the number of fields in a quantum
field theory}\ in combination with the exact Renormalization Group 
(RG) has the potential to be a very powerful \nonp\ technique.
The exact solution of the large $N$ limit along these lines offers
a tractable and possibly more general alternative to the traditional
approach of introducing collective (bound state) 
fields\ref\trad{see \eg  J. Zinn-Justin, 
            ``Quantum Field Theory and Critical Phenomena'' (1993)
             Clarendon Press, Oxford, or
S. Coleman, ``1/N'', Erice lectures (1979).}. Indeed, as we will show,
the exact solution of the large $N$ limit
 of the RG flow equations
does not require the introduction of
collective fields,
and yields solutions of a more general type than 
have been considered up until now. 

We will show how to solve exactly for the functional dependence of the
{\sl full} effective action in the large $N$ limit. However, of particular
interest is the flow equation for the potential, which closes in the
large $N$ limit, under certain conditions, and is closely related to the 
Local Potential Approximation (LPA).
A number of interesting results concerning the LPA
and this connection, have already been derived in refs.\ref\wegho{F.J. Wegner 
and A. Houghton, Phys. Rev. A8 (1973) 401.}\nref\ma{S.-K. Ma, Rev. 
Mod. Phys. 45 (1973) 589.}\nref\vved{D.D. Vvedensky, J. Phys.
A17 (1984) L251; A20 (1987) L197.}\nref\nicrev{T.S. Chang, 
D.D. Vvedensky and J.F. Nicoll, 
Phys. Rep. 217 (1992) 279}\nref\tetlit{N. Tetradis and D.F. Litim, Nucl. Phys. 
B464 (1996) 492.}\nref\Reu{M. Reuter, N. Tetradis and C. Wetterich,
Nucl. Phys. B401 (1993) 567.}\nref\tetwet{N. Tetradis and C. Wetterich, Nucl.
Phys. B422 (1994) 541.}\nref\Ell{U. Ellwanger and L. Vergara, Nucl. Phys.
B398 (1993) 52.}\nref\aoki{K-I Aoki {\it et al}, 
Prog. Theor. Phys. 95 (1996) 409,
hep-ph/9612458.}--\ref\com{J. Comellas and A. Travesset,  Barcelona preprint
UB-ECM-PF 96/21, hep-th/9701028.}, however, to our knowledge there is
no derivation of 
the general conditions, over and above the large $N$ limit,
 under which sufficient
simplifications occur to allow exact solution of the RG equations,
and/or such that these simplifications coincide with
the LPA. The main point of the present paper is to derive these conditions.


Specifically, we will show that the flow equations for $N$ component
 $O(N)$ invariant
scalar field theory simplify if and only if the one-particle irreducible
part of the interaction part of the effective action, \aka $\Gamma_\Lambda$
(to be defined below), is a functional only of the square of the field 
$\phi^2(\x)=\phi_a(\x)\phi_a(\x)$ at
some point $\Lambda$. Under rather general conditions, as discussed later, 
this occurs if the interaction part of
the bare action, $S_{\Lambda_0}$,
 is the (generalised) Legendre transform\ref\erg{T.R. Morris, 
Int. J. Mod. Phys. A9 (1994) 2411.}\
of such a Legendre effective action, \viz 
$\Gamma_{\Lambda_0}\equiv\Gamma_{\Lambda_0}[\phi^2]$. This space of
bare actions is indeed much larger than the one considered in traditional
large $N$ methods\trad, and in derivations of the large $N$ flow 
equations that start from these methods\ma\nicrev\Reu\aoki, where the
bare interactions were restricted to a general bare potential.

Although we concentrate on $O(N)$ scalar field theory,
it will be clear that the derivations (and
proofs) easily generalise to other field theories.

It is worth stressing that the large $N$ limit 
of the Legendre effective action {\sl does not} reduce simply to the sum of
a potential and a standard kinetic term. On the contrary, the Legendre
effective action generally
contains contributions to all orders in 
derivatives even in this large $N$ limit (as we will 
see later\foot{$O(\partial^2)$ results 
for general $N$, including $N=\infty$
\ref\mike{M. Turner and T.R. Morris, SHEP 97-06, hep-th/9704202},
were announced in ref.\ref\revii{T.R. Morris, in {\it RG96}, 
SHEP 96-25, hep-th/9610012.}.}).
Since the approximation in the LPA\ref\lpa{
J.F. Nicoll, T.S. Chang and H.E. Stanley, 
           Phys. Rev. Lett. 33 (1974) 540.}\ref\trunc{T.R. 
Morris, Phys. Lett. B334 (1994) 355.}\ref\truncm{T.R. 
Morris, Nucl. Phys. B458[FS] (1996) 477.}\
is the neglect of all these higher order terms, 
the LPA in the large $N$ limit  is even in this case
strictly speaking incorrect.  
However, the proof of exact reduction 
of the Legendre flow equation to that for just an effective potential
in the large $N$ limit, arises by considering only constant
fields (\ie ones that do not vary in space-time) and as we will see,
 so far
the reduced equations {\sl coincide} with the equations obtained in the LPA.

We will show that
the correct  large $N$ limit of the Wilson flow 
equations\ref\kogwil{K. Wilson and J. Kogut, Phys. Rep. 12C (1974) 75.}, or
equivalently\ref\deriv{T.R. Morris, Phys. Lett. B329 (1994) 241.}\ 
the large $N$ limit of Polchinski's flow equations\ref\pol{J. 
Polchinski, Nucl. Phys. B231 (1984) 269.}\ for the effective potential,
does not at all coincide with
the LPA. We will show that the 
 flow equations for the potential do again close in the
large $N$ limit however, and we derive the correct equations 
through a generalised Legendre transform relationship
between the Polchinski effective potential and the
 Legendre effective potential, which we formulate here. 
The generalised Legendre transform relationship, first written down
by one of us in ref.\erg, is the crucial step in this derivation.
The fact that the LPA is not the correct large $N$ limit in this case
 is remarkable, given that
the LPA of the Polchinski equation nevertheless provides the right 
eigenvalue spectrum for the Wilson-Fisher fixed point and the correct
behaviour for the BMB\ref\BMB{W.A. Bardeen, M. Moshe and M. Bander,
Phys. Rev. Lett. 52 (1984) 1188\semi F. David {\it et al}, 
Phys. Rev. Lett. 53 (1984) 2071, Nucl. Phys. B257 (1985) 695.}\
 phenomenon \com. However on the one hand,
we list several instances
in which the LPA of the Polchinski equation gives incorrect results,
and on the other hand, we
 show that flow equations of a very general form (of which the LPA is
one instance and the correct equation another)  all yield the
correct results for the eigenvalue spectrum of the 
Gaussian fixed point, the line of Gaussian fixed points that appear 
at $N=\infty$ in upper critical dimensions less than four\com\BMB,
and the  Wilson-Fisher fixed point.
In so doing, we also show that all the present forms of the RG can
be straightforwardly solved for the  corresponding RG eigenvalues. 

We finish this introduction with a brief discussion of
earlier work on the large $N$ limit of the RG\wegho-\com, 
in order to further clarify its relation to
the present paper.
In the classic paper of Wegner and Houghton\wegho, it was
already recognized that the effective potential for the
Wilson-Fisher fixed point,  
and its perturbations, were 
exactly soluble.
They noted that the functional flow equation
could be exactly
reduced to a partial differential equation for the 
effective potential.
Later it was noticed that 
all points on the flow
were soluble, for any bare potential
in this so-called LPA\vved\ (see also refs.\nicrev\tetlit\com). 
Actually, the exact solution was already published and analysed by
Ma\ma, who solved for any finite renormalization group step 
 for the effective
potential by now-standard large $N$ methods.
These  exact results were all derived for the potential only and for
sharp cutoff. 
Apart from a solution for a particular smooth cutoff\mike,
we know of no other published
exact results\foot{In ref.\tetwet\ the large $N$ equation was further
approximated and solved for. The authors speculate
that this approximation may become exact for 
a certain choice of cutoff function. However,
this is not correct since the approximated integral 
$L^3_1(u')$\tetwet\ has a cut from $u'=-1$ for all sensible cutoffs, not
a double pole at $u'=-1$ as the authors claimed.}\
 extracted from the large $N$ flow
equations. As we have already stated,
 we will show in fact that it is possible to go
well beyond this: that all the
equations are soluble {\sl for any cutoff}, both around the Wilson-Fisher
fixed point and general points, and indeed not just for the potential 
but for the full effective action.

In ref.\aoki,
the expansion of the potential about its minimum was investigated
and shown to yield a partial system of `perfect 
coordinates' for the RG. 
This property of perfect coordinates leads to an
elegant derivation of the RG eigenvalues, 
which we will utilise later, for
we will show that this property 
 holds not just for sharp cutoff (\viz the Wegner-Houghton RG\wegho)
but for all cutoffs and all
present forms of the RG.
By thus showing that truncations of the Taylor expanded flow equations
around the minimum of the potential, become exact 
in the limit $N\to\infty$, we also gain some
insight into why
truncations around the minimum give 
good results in finite 
$N$ scalar field theory \tetwet\ref\alford{M. Alford, 
Phys. Lett. B336 (1994) 237.}.

\noindent{\bf 2. The large $N$ limit of the Legendre flow equation.}

As stated earlier, for concreteness 
we will restrict our attention to an $N$-component
$O(N)$ invariant scalar
field theory. However, the appropriate generalizations to the other 
classical groups, and other (matter) fields is straightforward.

Consider the partition function, with kinetic term
 modified to include an infrared cutoff:
\eqn\zorig{\exp {\cal W}[J]=\int\!\D\phi\
\exp\{-\half\phi_a\cdot\Delta_{IR}^{-1}\cdot\phi_a-S_{\Lambda_0}[\phi]
+J_a\cdot\phi_a\}\quad,} 
We have introduced $\Delta_{IR}(q,\Lambda)= C_{IR}(q^2/\Lambda^2)/ q^2$, 
where $C_{IR}$ is the infrared cutoff function.
The subscript ${}_a$ is the internal index running from 1 to $N$, 
and $S_{\Lambda_0}$ stands
for the $O(N)$ invariant bare interactions.
Some more or less pedantic statements, on notation, and $C_{IR}$, are
given in appendix A. 
Differentiating \zorig\ with respect to $\Lambda$ and
rewriting in terms of $\Gamma_\Lambda$, the 
interaction part of the Legendre effective action,
gives\erg\deriv\ref\nici{J.F. Nicoll and 
T.S. Chang, Phys. Lett. 62A 
(1977) 287;\ C. Wetterich, Phys. Lett. B301 (1993) 90;\
M. Bonini {\it et al},
Nucl. Phys. B409 (1993) 441.}:\eqna\leg\
$$\eqalignno{
{\partial\over\partial\Lambda}\Gamma_\Lambda[\phi] &=
-{1\over2}\,\tr\ {1\over \Delta_{IR}}{\partial 
\Delta_{IR}\over \partial\Lambda}\cdot \left(A^{-1}\right)_{aa} &\leg a\cr
\ins01{where} A_{ab} &=\delta_{ab}+\Delta_{IR}\cdot{
\delta^2\Gamma_\Lambda\over\delta\phi_a\delta\phi_b}\quad. &\leg b\cr}$$
Here $\Gamma_\Lambda$ is defined by
$\Gamma_\Lambda[\phi]+\half\phi_a\cdot\Delta_{IR}^{-1}
\cdot\phi_a=-{\cal W}[J]+J_a\cdot\phi_a$,
where now $\phi_a=\delta {\cal W}/\delta J_a$ is the classical field.

Suppose that the interaction part of the
Legendre effective action is just a functional
of $\phi_a(\x)\phi_a(\x)$, \ie $\Gamma_\Lambda=
\Gamma_\Lambda[\phi^2]$  (at least at some value of $\Lambda$), then we have
\eqn\ed{{\delta^2\Gamma_\Lambda\over\delta\phi_a(\x)\delta\phi_b(\y)}
=2\delta_{ab}\delta(\x-\y){\delta\Gamma_\Lambda\over\delta\phi^2(\y)}
+4\phi_a(\x)\phi_b(\y){\delta^2
\Gamma_\Lambda\over\delta\phi^2(\x)\delta\phi^2(\y)}
\quad.}
We note that this is of the form $a\delta_{ab} +b_{ab}$.
If $\Gamma_\Lambda$ has a smooth limit as $N\to\infty$, 
then it is easy to see
that the second term $b_{ab}$ contributes
negligably in this limit. Indeed, whereas the other terms in \ed\
and \leg{b}\ yield a factor of $\delta_{aa}=N$ on taking the inverse
and tracing in \leg{a}, corrections resulting from $b_{ab}$
have no $\delta_{aa}$ component to all orders in $b$:
$(a\delta_{ab}+b_{ab})^{-1}=a^{-1}\delta_{ab}-a^{-1}b_{ab}a^{-1}
+a^{-1}b_{ac}a^{-1}b_{cb}a^{-1}-\cdots$.
Therefore to leading
order in $N$,
equation \leg{}\ simplies to \eqna\lege\
$$\eqalignno{{\partial\over\partial\Lambda}\Gamma_\Lambda[\phi] &=
-{N\over2}\,\tr\ {1\over \Delta_{IR}}{\partial 
\Delta_{IR}\over \partial\Lambda}\cdot A^{-1} &\lege a\cr
\ins01{where now} A(\x,\y) &= \delta(\x-\y) 
+2\Delta_{IR}(\x-\y){\delta\Gamma_\Lambda\over\delta\phi^2(\y)}
&\lege b\cr}$$
(the propagator $\Delta_{IR}$ having been Fourier transformed into position
space). Two points are now obvious from \lege{}. 
Firstly, 
if for some value $\Lambda$,
$\Gamma_\Lambda$ is a functional only of $\phi^2$, then since \lege{b}\
is also only a functional of $\phi^2$, and \lege{a} a first order 
differential equation in $\Lambda$,  
we have 
immediately that this form of $\Gamma_\Lambda$ is preserved by the flow, 
\ie $\Gamma_\Lambda\equiv
\Gamma_\Lambda[\phi^2]$ for all $\Lambda$. 
Secondly, in order to obtain a finite non-trivial limit as $N\to\infty$,
we must transform 
\eqn\trans{\Gamma_\Lambda\mapsto N\,\Gamma_\Lambda, \ins11{and consequently} 
\phi_a\mapsto \sqrt{N}\,\phi_a\quad.}
 Of course these are nothing but the
standard transformations\trad, but we have here yet another derivation
of them. The net effect of this is to remove the factor of $N$ from \lege{a}.
We assume from now on that this has been done.

The fact that the flow equation \lege{}\ now contains only zero or one
point functions allows for a considerable simplification: Set $\phi^2(\x)=z$,
where $z$ is a constant. Regarding the effective action as given by
a derivative expansion (or otherwise), we see that only the potential term
of $\Gamma_\Lambda=\int\!d^D\!x\ V(\phi^2,\Lambda) +\cdots$
survives in \lege{},  which thus takes the form 
\eqn\Vfl{
{\partial\over\partial\Lambda} V(z,\Lambda)
=-{\Omega\over2}\int_0^\infty 
\!\!\!\!dq\, q^{D-1}{1\over \Delta_{IR}}{\partial 
\Delta_{IR}\over \partial\Lambda}{1\over 1+2\Delta_{IR}
V'(z,\Lambda)} }
(where in here $\Delta_{IR}\equiv \Delta_{IR}(q,\Lambda)$,
 prime refers to differentiation with respect to 
$z$, and $\Omega (2\pi)^D$ is the solid angle of a $(D-1)$-sphere). 

Evidently \Vfl\  coincides with what would be
obtained in the Local Potential Approximation,
since this results from the ansatz 
 $\Gamma_\Lambda=\int\!d^D\!x\, V(\phi^2,\Lambda)$, combined with the
requirement that all  higher derivative terms are discarded.
Nevertheless contributions from all higher derivatives are most  certainly
(generally) there, as is evident from the full equation \lege{}.

The conditions under which this `reduction to LPA' takes place, follow from the
first point made below \lege{}, \ie a sufficient condition
is that $\Gamma_\Lambda=
\Gamma_\Lambda[\phi^2]$ at some point $\Lambda$.
In particular, if the ultraviolet
cutoff is absorbed in $C_{IR}$, by modifying it so that \eg 
$C_{IR}(q,\Lambda)=0$ for all $q>\Lambda_0$ then we have that
$\Gamma_{\Lambda_0}=S_{\Lambda_0}$ \erg, and thus the reduction occurs
if the bare interactions are functionals only of $\phi^2$.
 (If the ultraviolet cutoff is not so absorbed 
in $C_{IR}$, then generally $\Gamma_\Lambda\to S_{\Lambda_0}$ as 
$\Lambda\to\infty$ and the conclusions will be the same.)
 One can convince
oneself that the condition $\Gamma_\Lambda=
\Gamma_\Lambda[\phi^2]$ at some point $\Lambda$,
 is also necessary: If $\Gamma_\Lambda$ contains
a term of the form \eqn\cou{\int\!d^D\!x\, f(\phi^2)\,\phi_a B\, \phi_a\quad,}
where $f$ is some function and $B\equiv B(-i\partial)$ a two-point function
satisfying $B(0)=0$, here
expressed as a derivative operator, then \ed\ receives corrections
containing $\delta_{ab}$. These corrections thus survive to \lege{},
and furthermore, 
the $\delta_{ab}$ term arising from differentiating the two fields in
$\phi_a B\phi_a$, modifies \Vfl\
in way which can only be quantified once $B$ and $f$ are known.
It is only contributions to the potential, and
terms (or parts of terms) that can be 
put in form \cou\
that survive the twin reductions of the large $N$ limit and setting
the field to a constant. This fact however does not appear to allow any
widening of the above condition, such that the equations remain
soluble, because
terms of the form \cou\ receive their corrections through
the flow equation from many other forms of terms
and no closure appears possible in general. In particular therefore, it
appears that no closure allows the computation of a non-trivial $f$ and $B$
in general, nor does a closure exist which guarantees that these terms vanish,
beyond the already stated
sufficient condition.

\noindent{\bf 3. The LPA of Polchinski's flow equation.}

Consider now Polchinski's flow equation for the interaction part 
$S_\Lambda[\Phi]$ of the Wilsonian effective action
$S^{eff}_\Lambda=
\half\Phi_a.\Delta_{UV}^{-1}.\Phi_a +S_\Lambda$.
This is given by\pol\erg
\eqn\Pol{
{\partial S_\Lambda \over\partial\Lambda}={1\over2}\,
 {\delta S_\Lambda\over\delta\Phi_a}\cdot
{\partial\Delta_{UV}\over\partial\Lambda}\cdot
{\delta S_\Lambda\over\delta\Phi_a}
-{1\over2}\, {\rm tr}\, 
{\partial\Delta_{UV}\over\partial\Lambda}\cdot
{\delta^2S_\Lambda\over\delta\Phi_a\delta\Phi_a}
\quad.}
$\Delta_{UV}(q,\Lambda)=C_{UV}(q,\Lambda)/q^2$, 
and $C_{UV}$ is an effective ultra-violet
cutoff function. Its properties (and that of $\Phi$) are recalled
in appendix A. We remind the reader that Wilson's flow equation\kogwil\
is identical to \Pol, after the transformation 
$\Phi_a\mapsto\sqrt{C_{UV}}\,\Phi_a$ (and 
${\cal H}\equiv- S_\Lambda$) \deriv\revii, and therefore does not
need separate discussion. Similarly, the Wegner-Houghton equation arises as
the sharp cutoff limit of eqn.\Pol\ \erg\revii.

A finite large $N$ limit is achieved with the analagous 
transformations to \trans, \viz 
\eqn\ptrans{S_\Lambda\mapsto N\, S_\Lambda\quad,\hskip1cm \Phi_a
\mapsto \sqrt{N}\,\Phi_a\quad.}
[Again these either follow from standard
methods, 
or may be deduced directly, this time by requiring the order
$N$ parts of the quantum (trace) term to balance the classical 
($\sim S^2_\Lambda$) term and the left hand side in eqn.\Pol.]
Suppose once more that we assume that the interactions are restricted
to the form $S_\Lambda= S_\Lambda[\Phi^2]$, at some value
of $\Lambda$. Combining this with
\ptrans, and $N\to\infty$, we obtain \eqna\Polch
$$\eqalignno{
{\partial\over\partial\Lambda}S_\Lambda[\Phi] &=
\int\!\!d^D\!x \left\{
2\left({\delta S_\Lambda\over\delta\Phi^2}\Phi_a\right)\,
{\partial\Delta_{UV}\over\partial\Lambda}(-i\partial,\Lambda)\,
\left(\Phi_a
{\delta S_\Lambda\over\delta\Phi^2}\right)
-\gamma\Lambda^{D-3}\,{\delta S_\Lambda\over\delta\Phi^2}\right\}
\ \ &\Polch a\cr
\ins01{where} \gamma &=\Omega\Lambda^{3-D} \int^\infty_0\!\!\!\!dq\, q^{D-1}
{\partial\Delta_{UV}\over\partial\Lambda}(q,\Lambda) \quad.  &\Polch b\cr}
$$

Because in this limit, the only field derivatives are first order, we 
have again a reduction to LPA on setting $\Phi_a(\x)$ constant,
 $\Phi^2(\x)=y$ (not to be confused with
the coordinate \y [!]): \eqnn\lpaP\ \eqnn\alph
$$\eqalignno{
{\partial\over\partial\Lambda}U(y,\Lambda) 
&={4\alpha\over\Lambda^3}
y \left(U'\right)^2 -\gamma\Lambda^{D-3} U'\quad, &\lpaP \cr
\ins01{where} \alpha &=-C'_{UV}(0) &\alph \cr}$$
(the prime on $C_{UV}$ meaning differentiation with respect to its
argument). Nevertheless, eqn.\lpaP\ is not correct (for all $\Lambda$).
This is because the form $S_\Lambda=S_\Lambda[\Phi^2]$ is violated by
the first term on the right hand side of \Polch{a}\ and is thus not
preserved by the flow. 
Independently, there are several ways to see that \lpaP\ cannot be
the exact equation. Firstly, it fails to give the correct 
$\beta$ function in $D=4$ dimensions\revii\ref\conv{T.R. Morris,
in preparation.}. Secondly, there is nothing wrong with a cutoff
function satisfying $C_{UV}'(0)=0$, but \lpaP\ collapses to a
linear equation which cannot be right since in particular it cannot
support non-trivial fixed points.\foot{In fact it is straightforward
to solve the $\alpha=0$ case and confirm its trivial behaviour.}\
Indeed the sharp cutoff limit results in $\alpha=0$, where 
the exact equation should collapse to the LPA of Wegner-Houghton
equation \lpa\trunc\truncm.
Finally, a general Legendre transform relationship exists between
the Polchinski and Legendre effective 
actions\erg\truncm\ref\eqs{T.R. Morris, Southampton preprint 96-36,
hep-th/9612117.}\ which implies that precisely when
$\alpha=0$, the Polchinski and Legendre effective potentials
should be identical. 

\noindent{\bf 4. Legendre transform relations.}

 Here, we develop the consequences of the Legendre transform
relationship for the two
effective potentials. The content of this
section holds generally, \viz not just for large $N$, with 
the exception of the scaling 
form for the cutoffs which, for
 smooth cutoffs, is only appropriate for fixed points and fields
with no anomalous scaling (\cf refs.\deriv\truncm\revii\
 and appendix A) -- the
generalization is straightforward however. 
  
The Polchinski (or Wilson) effective action has a 
tree structure\pol\erg\ whose one-particle irreducible parts are 
given by the infrared cutoff Legendre effective action introduced
previously \erg, providing that we set $C_{IR}(q^2/\Lambda^2)=1-
C_{UV}(q^2/\Lambda^2)$. The interactions 
are then related by the Legendre transform relationship\erg\
\eqn\legtr{S_\Lambda[\Phi]=\Gamma_\Lambda[\phi]+\half
(\phi_a-\Phi_a)\cdot\Delta_{IR}^{-1}\cdot(\phi_a-\Phi_a)\quad,}
which by differentiation, implies the following functional relations
between the two fields:\eqna\deph\
$$\eqalignno{
\phi_a &=\Phi_a-\Delta_{IR}\cdot {\delta S_\Lambda\over\delta\Phi_a}
 &\deph a\cr
\ins01{and} \Phi_a &= \phi_a +\Delta_{IR}\cdot
{\delta\Gamma_\Lambda\over\delta\phi_a} \quad. &\deph b\cr}$$
It is straightforward to show that the flow equations are transformed
into each other by \legtr;  By differentiating \deph{a}\
with respect to $\Phi$, and \deph{b}\ with respect to $\phi$, one
derives 
\eqn\twopt{\left(A^{-1}\right)_{ab}=\delta_{ab}-\Delta_{IR}\cdot
{\delta^2 S_\Lambda\over \delta\Phi_a\delta\Phi_b}\quad,}
then \eg differentiating \legtr\
with respect to $\Lambda$ at constant $\Phi$, and using \deph{a}\
and the above relation, one derives \Pol\  from \leg{} (up to 
an unimportant vacuum energy\pol\erg). 

In the case of constant
fields with $\Phi_a(\x)={\hat\Phi}_a\sqrt{y}$ and 
$\phi(\x)={\hat\phi}_a\sqrt{z}$
(where we have introduced the unit vectors ${\hat\Phi}^2={\hat\phi}^2=1$),
 \legtr\ collapses to 
\eqn\legtrUV{
U(y,\Lambda)=V(z,\Lambda)+{\Lambda^2\over2\alpha}(\phi-\Phi)^2\quad,}
and either directly from here or from \deph{}, we see that
${\hat\phi}_a={\hat\Phi}_a$,
$$\sqrt{{y\over z}} =1+{2\alpha\over\Lambda^2}V'(z,\Lambda)\ins11{and}     
\sqrt{{z\over y}}=1-{2\alpha\over\Lambda^2}U'(y,\Lambda)\quad.$$
Combining the above equations gives,
\eqn\UtoV{V'(z,\Lambda)={U'(y,\Lambda)\over1
-2\alpha U'(y,\Lambda)/\Lambda^2}\quad.}
We have in particular at 
$\alpha/\Lambda^2\equiv \Delta_{IR}(0)=0$, corresponding to an  
efficient infrared cutoff, that the constant fields and 
the  effective potentials are equal:
$y=z$, $U=V$ \erg\truncm. Consequently, 
the Polchinski effective potential tends to the full\foot{\ie without
infrared cutoff}\ Legendre effective
potential in the limit as $\Lambda\to0$ \eqs.  
For $\alpha>0$, \UtoV\ is compatible with the observed large field
behaviour of the respective potentials:  
we have 
$V'\sim {\tilde A} z^{2/(D-2)}$, where ${\tilde A}$ is a dimensionless
constant (by dimensions on assuming $\Lambda$
drops out of this limit) \deriv\eqs, and thus from \UtoV, 
$U'\sim {\Lambda^2\over2\alpha}-{\Lambda^4\over4\alpha^2{\tilde A}}
z^{2/(2-D)}$, concurring with derivative expansion 
estimates\ref\balls{R.D. Ball et al, Phys. Lett. B347 (1995) 80.}.

\noindent{\bf 5. The large $N$ limit of Polchinski's flow equation.}

Returning to \Polch{a}, we note that the form $S_\Lambda\equiv 
S_\Lambda[\Phi^2]$ is not preserved in the large $N$ limit,
 precisely because the 
one-particle reducible parts do not respect this form. 
The best one can derive for the full
 $S_\Lambda$, is its tree expansion in terms
of one-particle irreducible vertices, the latter simplifying as a
consequence of $\Gamma_\Lambda\equiv\Gamma_\Lambda[\phi^2]$, \eg
we derive  from \twopt\ and \lege{b}, at $\phi=\Phi=0$
the large $N$ form for the effective two-point function. 

However, for the large $N$ Polchinski effective potential 
we can, by using the results of the last section,
 derive an exact closed equation -- 
either directly from \legtrUV, \UtoV\ and \Vfl, along the lines 
outlined below \Vfl, or by setting $\Phi_a={\hat\phi}_a \sqrt{y}$,
in \Pol\ and using \twopt, \lege{b}\ and \UtoV. Either way, we
obtain
\eqn\Ufl{{\partial U\over\partial\Lambda} 
={4\alpha\over\Lambda^3} y \left(U'\right)^2
+{\Omega\over2}\left( {2\alpha\over\Lambda^2}U'-1\right)
\int^\infty_0\!\!\!\!dq\, q^{D-1}{1\over \Delta_{IR}}{\partial 
\Delta_{IR}\over \partial\Lambda}{1\over 1+2(\Delta_{IR}-\alpha/\Lambda^2)
U'}\quad.}

\noindent{\bf 6. Exact large $N$ solutions.}

Exact large $N$ solutions of RG effective potentials
have been found for: the sharp 
cutoff (Wegner-Houghton) case, around the Wilson-Fisher 
fixed point\wegho\aoki\ and
at general points\nicrev\tetlit\com,
the Legendre flow equation with a particular smooth cutoff --
around the Wilson-Fisher fixed point\mike, and the large $N$ LPA 
\lpaP\ for
Polchinski's flow equation at general points\com. 
In this section, we show that these equations,
and \Ufl, are exactly soluble {\sl for any cutoff},
both around the Wilson-Fisher fixed point and at general points, 
 and that this is possible not just for the potential,
but also for the full effective action $\Gamma_\Lambda[\phi^2]$.
At the same time, we will shed light on why \lpaP\ is capable of
producing the correct exact 
results\com\ despite the fact that \lpaP\ is not
itself exact. 

Since corrections are functionals of $\phi^2$ only, 
the kinetic term $\sim(\partial_\mu\phi)^2$ remains normalised,
and the fields have no anomalous dimension. Thus we write
$z\mapsto z\Lambda^{D-2}$,  
$V(z,\Lambda)\mapsto V(z,t) \Lambda^D$, $t=\ln(\mu/\Lambda)$
(where $\mu$ is an arbitrary finite energy scale),
and similarly for $y$ and $U$, after which all powers of $\Lambda$ may
be eliminated from \Vfl, \lpaP\ and \Ufl.
Shifting to a stationary point\foot{
{\it N.B.} By this we restrict ourselves to potentials with such 
stationary points, ruling out in particular the High Temperature
fixed point solution\com\ to \Pol: $U(y,t)=y/(2\alpha)$.}\ 
$x_0(t)$, by writing $z=x_0(t)+x$ [$y=x_0(t)+x$], we define
$W(x,t)=V'$ [$W(x,t)=U'$], with thus $W(0,t)\equiv0$. 
Differentiating 
\Vfl, \lpaP,  and \Ufl\ by $x$, all three flow
equations can then be written 
(for general cutoff) in the form
\eqn\allfl{\dot W-\dot{x}_0 W'+(D-2)(x_0+x)W'-2W=\gamma W'
-\left(d[x+x_0]+c-{d\gamma\over D-2}\right)WW'+F\quad,}
where  the dot $\dot{}$ stands for 
$\partial/\partial t$ (at constant $x$),
and $F(W,W',x,t)$ is an analytic function of $W$ (around $W=0$), 
$W'$ and $x$, of $O(W^2)$.   
$\gamma$, $d$ and $c$ (chosen in the above form for 
convenience), are numerical constants.  

We now prove that all flow 
equations of form \allfl, yield the correct structure of RG fixed points
in the sense that: There exists a Gaussian fixed point $W=0$, 
with RG eigenvalues $\lambda^G_n=2-(D-2)n$, which 
becomes a line of Gaussian fixed points in the upper critical
dimensions $D_k=2+2/k$, $k=2,3,\cdots$ \BMB\com. There exists
a non-trivial (Wilson-Fisher -- or Heisenberg) fixed point 
with eigenvalue spectrum
$\lambda^{WF}_n=D-2(n+1)$, $n=0,1,\cdots$ \wegho,
--  providing only that $c\ne0$.
Furthermore, if $c$ (for $D>2$) and $\gamma$ are positive, 
which we will assume, then
 the Wilson-Fisher
fixed point derived from \allfl\ exists, as expected,
 only in dimensions $2<D<4$.
For our three flow equations, $\gamma=2\Omega\int\!\!dq\,q^{D-1} C'_{IR}(q^2)$,
as in \Polch{b}. $d=8\alpha$ for the Polchinski types \lpaP, \Ufl, while
$d=0$ for \Vfl. $c=c_L\equiv8\Omega\int\!\!dq\,q^{D-3}C'_{IR}C_{IR}$
for \Vfl, $c=c_P\equiv 8\alpha\gamma/(D-2)$ for \lpaP, and
$c=c_L+c_P-4\alpha\gamma$ for \Ufl. We note that for $C_{IR}$ monotonic,
$\gamma$, $c_L$ and (for $D>2$) $c_P$ 
are guaranteed positive (\cf appendix A), and \Ufl's $c$  is
positive at least in the range $2<D<4$. 
 
Requiring $W(0,t)=0$ in \allfl, implies $W'(0,t)=0$
and thus\foot{by expanding in powers of $x$ or otherwise}
 $W(x,t)=0$, unless
\eqn\xfl{\dot{x}_0=(D-2)x_0-\gamma\quad.}
Substituting this back into \allfl, and
expanding  $W(x,t)=\sum_{n=1} W_n(t) x^n$, we obtain
\eqnn\wi\eqnn\wn
$$\eqalignno{
\dot W_1+(D-4)W_1 &=-\left[dx_0+c-{d\gamma\over D-2}\right] W_1^2\quad, &\wi\cr
\dot W_n+\left[(D-2)n-2\right]W_n &=-\left[dx_0+c-{d\gamma\over D-2}\right]
(n+1)W_nW_1 +\cdots\quad, &\wn\cr}$$
where the second equation holds for $n>1$ and the dots stand for terms
containing products of $W_m$'s,  $m<n$.
Thus the $W_n(t)$ are all soluble in terms of lower $m$ $W_m(t)$ and $x_0(t)$,
and truncations
to some finite $n$ are exact. This extends 
ref.\aoki's observation of `perfect coordinates',
to all cutoffs and flow equations.
Consider now a fixed point: $W(x,t)=W(x)$, $x_0(t)=x_0$.
 From \xfl, $x_0=\gamma/(D-2)$ which,
since $x_0$ is the original field squared, requires $D>2$. 
Substituting into \wi, we find that
either $W_1=0$ -- which is discussed further below,
or $W_1=(4-D)/c$ -- which is the solution we take here. Since
this implies that $W(x)$ crosses the $x$-axis only once, 
we must have $W_1>0$, and thus $D<4$, so that $W(x)>0$ for $x>0$.  
Otherwise, from the large field
behaviours below \UtoV, 
the potential (if it exists)\foot{In fact such potentials typically
encounter a singularity at some point\revii\ref\hh{
T.R. Morris, Phys. Rev. Lett. 77 (1996) 1658.}\com.}\
 is unbounded below.  
Substituting these solutions 
in \wn, we see that the $W_n$ exist and are
unique, with no further conditions.
Now, from \xfl, we see that about this fixed point,
there is an eigen-perturbation  with 
$\delta x_0(t)\propto{\rm e}^{\lambda^{WF}_0 t}$. 
From \wi\ and \wn, 
there is just one new eigen-perturbation 
for each new $n=1,2,\cdots$, and this takes the form  $\delta x_0(t)=0$, 
$\delta W_m(t)=0$ $\forall m<n$,
and $\delta W_n(t)\propto{\rm e}^{\lambda^{WF}_n t}$.
This completes the space of perturbations, and the 
eigenvalue spectrum of  the Wilson-Fisher fixed point.

Now we return to the fixed point solution $W_1=0$. It is easy to see
(by induction in increasing $n$) that at generic dimension $D$,
\wn\ implies $W_n=0$ $\forall n$,
 \ie the Gaussian fixed point.
Since $W_1=0$, the eigenvalues of the perturbations are just given by
the coefficient on the left hand sides of \wi\ and \wn:
$\lambda_n=\lambda^G_n$. 
But at $D=D_k$, the coefficient of $W_k$ for $n=k$ in \wn, vanishes.
Therefore in this case $W_k$ can be arbitrary.
Since $W_1=0$, the eigenspectrum remains that of the Gaussian
fixed point; $W_k$ is an exactly marginal coupling and the fixed point
has become a line. ({\it N.B.} Since the coefficient of the $x$
terms in \allfl\ vanish as a consequence of $W_1=0$, \xfl\ does not
yield an eigenperturbation. The mass perturbation,
$\delta W(x,t)\propto {\rm e}^{\lambda^G_0t}$,
on the other hand, exists but was excluded by our condition
$W(0,t)=0$.)

Note that these proofs do not provide closed forms for 
non-trivial $W(x)$, or establish
that they are non-singular for all $x+x_0>0$ (and thus in particular
the extent of the line of Gaussian fixed points at $D=D_k$)
nor similarly, that the perturbations $\delta W(x,t)$
are acceptable\revii\hh. This can be established from general properties
of the equations,
similarly to the proofs in refs.\wegho\ref\ui{T.R. Morris, 
Phys. Lett. B357 (1995) 225.}, or by deriving exact implicit solutions. 
Indeed these are easily achieved for 
all initial conditions and all $\Lambda$: For simplicity
restricting to the Legendre
flow equation and before scaling,\foot{It is straightforward 
to generalise, -- or  use \UtoV\ if necessary, and scale $\Lambda$
out afterwards.}\  we 
differentiate \Vfl\ with respect to $z$ and write 
$W(z,\Lambda)=V'(z,\Lambda)$,
after which \Vfl\ takes the
form $\partial W/\partial\Lambda ={\cal A}(W,\Lambda)\, W'$. Now by
change of dependent variables (or the method of characteristics), the
general solution is 
$$z= {\cal F}\left(W(z,\Lambda)\right)
-\int^\Lambda\!\!\!\!d\chi\ {\cal A}\left(W(z,\Lambda),\chi\right)\quad,$$
where ${\cal F}$ is an arbitrary function. A closely related technique
works for the full effective action $\Gamma_\Lambda[\phi^2]$. Let us
define a Legendre transform with respect to $\phi^2(\x)$, introducing
a source $K(\x)$. Writing ${\cal W}[K]=-\Gamma_\Lambda[\phi^2]+K\cdot\phi^2$,
we derive in standard fashion ${\delta\Gamma_\Lambda\over\delta\phi^2}=K$,
and thus using \lege{}\ we obtain an equation for ${\partial{\cal W}\over
\partial\Lambda}|_K$ whose general solution is immediate:
$${\cal W}= {\cal W}_0[K]+{1\over2} \int^\Lambda\!\!\!\!d\chi\ \tr\ 
{1\over \Delta_{IR}}{\partial 
\Delta_{IR}\over \partial\chi}\cdot A^{-1}\quad,$$
where ${\cal W}_0$ is an arbitrary $\Lambda$ independent functional of $K$,
 $\Delta_{IR}\equiv\Delta_{IR}(q,\chi)$ and from \lege{b}, 
$A(\x,\y)=\delta(\x-\y)+2\Delta_{IR}(\x-\y)K(\y)$.

We finish with two remarks. Firstly, it looks quite plausible that the
`perfect coordinate' method \xfl--\wn\ 
can be generalised to the full functional
equation \lege{}. Secondly,  we are aware that a number of the results
we have given may
be interpreted and even derived at the diagrammatic level, however
the derivations we give are compact,  and non-perturbative,  and
are thus more powerful.

\bigbreak\bigskip\bigskip
\centerline{{\bf Acknowledgements}}\nobreak
TRM acknowledges support of 
 the SERC/PPARC through an Advanced Fellowship, and PPARC grant
GR/K55738.

\vfill\eject

\appendix{A}{Notation and cutoffs.}
We use a condensed
notation wherever convenient
so two-point functions are often regarded as matrices in position or 
momentum ($\q$)
space, one-point functions as vectors, and contractions indicated by a
dot. We work in $D$ dimensions. The partition function is
assumed to be regulated by an overall cutoff $\Lambda_0$,
but all equations of 
interest will be regulated by the intermediate cutoff $\Lambda$, and
are
thus completely insensitive to $\Lambda_0$ in the continuum limit
$\Lambda_0\to\infty$. For more details see refs.\erg\deriv\trunc\truncm.

A more general propagator than $1/q^2$ could be used (as multiplier to
$C_{IR}$ and $C_{UV}$)  without any significant changes, but at the
price of some loss of clarity. 
Similarly the cutoff functions could be assumed to be general functions
$C(q,\Lambda)$ of yet to be determined dimension, but since in the
large $N$ limit for scalar fields no anomalous dimension remains, and
since we will later in the paper be interested in  (scale-free) 
critical points we simplify from the start
to $C\equiv C(q^2/\Lambda^2)$. 

The infrared cutoff function $C_{IR}$
 satisfies
$C_{IR}(r)\to1$ as $r\equiv q^2/\Lambda^2 \to\infty$,
so that physics is independent of the infrared cutoff at scales
much larger than $\Lambda$. In fact we require also that 
$C_{IR}(r)\to1$ sufficiently fast that the
momentum integrals in the effective action
flow equations  \leg{}\
are well regulated.
For $r\to0$, we require  $C_{IR}(r)\to0$,
so that the infrared physics is indeed cutoff.
In particular we will require that this happens fast enough to guarantee
that a momentum Taylor expansion (\aka derivative expansion) 
exists
for the effective action (\ie that there are no remaining massless modes).
By unitarity, the cutoff  functions must be analytic at $r=0$,
(so that they hide no other massless modes).

The effective ultraviolet cutoff function $C_{UV}(r)$ is required
to satisfy  $C_{UV}\to1$ as $r\to0$,
so that physics is unchanged
 at scales much less than the effective cutoff $\Lambda$,
while for $r\to\infty$, $C_{UV}$ is required to vanish
sufficiently fast that all momentum integrals are well regulated.
In fact, we will insist that $C_{UV}(r)+C_{IR}(r)=1$ for all $r$,
since then an intimate relation exists\erg\ 
between the Polchinski and Legendre
flow equations. 
The field $\Phi=\Delta_{IR}\cdot J+\phi$ is not the same
as the field $\phi$ in \zorig\ \erg, unless Polchinski's 
conditions\pol\ are applied to $C_{UV}$ and $J$. These
conditions are not assumed, nor necessary, here.

\listrefs

\end